\documentclass[aps,prd,nofootinbib,superscriptaddress,amsmath,amssymb,twocolumn]{revtex4}
\usepackage{graphicx,epsf,float}
\usepackage[]{latexsym}
\usepackage[utf8]{inputenc}

\newenvironment{seqn}{\equation\aligned}{\endaligned\endequation}
\newcommand{\be}{\begin{seqn}}
\newcommand{\ee}{\end{seqn}}
\newcommand{\bea}{\begin{eqnarray}}
\newcommand{\eea}{\end{eqnarray}}

\def\comment#1{}

\usepackage{color}

\definecolor{darkred}{rgb}{.8,0,0}

\definecolor{darkblue}{rgb}{0,0,.7}

\definecolor{darkgreen}{rgb}{0,.7,0}

\begin{document}

%
%
\title{Proton decay and the quantum structure of spacetime}

%

%
%
%
%
\author{Abeer~Al-Modlej}
\email{amodlej@ksu.edu.sa}
\affiliation{ Department of Physics and Astronomy, King Saud University, Riyadh 11451, Saudi Arabia}
\author{Salwa~Alsaleh}
\email{salwams@ksu.edu.sa}
\affiliation{ Department of Physics and Astronomy, King Saud University, Riyadh 11451, Saudi Arabia}
\author{Hassan~Alshal}
\email{halshal@sci.cu.edu.eg}
\affiliation{Department of Physics, University of Miami, Coral Gables, FL 33124, USA}
\affiliation{Department of Physics, Faculty of Science, Cairo University, Giza 12613, Egypt} 
\author{Ahmed~Farag~Ali}
\email{Ahmed.ali@fsc.bu.edu.eg}
\affiliation{Department of Physics, Faculty of Science, Benha University, Benha, 13518, Egypt}
\affiliation{Quantum Gravity Research, Los Angeles, CA 90290, USA}

\date{\today}
\begin{abstract}
\begin{center}
\textbf{Abstract}
\end{center}
Virtual black holes in noncommutative spacetime are investigated using coordinate coherent state formalism such that the event horizon of black hole is manipulated by smearing it with a Gaussian of width $ \sqrt \theta $, where $ \theta$ is the noncommutativity parameter. Proton lifetime, the main associated phenomenology of the noncommutative virtual black holes, has been studied: first in $4$ dimensional spacetime and then generalized to $D$ dimensions. The lifetime depends on $ \theta$ and the number of spacetime dimensions such that it emphasizes on the measurement of proton lifetime as a potential probe for the micro-structure of spacetime.
\end{abstract}
\maketitle

\vspace{-0.75cm}
\section*{Introduction}
Since the introduction of Hawking radiation \cite{Hawking:1974sw}, many semiclassical, non-classical \cite{Birrell:1982ix,DeWitt:2003pm,Fulling:1989nb,Wald:1995yp,Ford:1997hb,Bar:2009zzb,Parker:2009uva,Donoghue:2017pgk} and \emph{geometrical} \cite{Sardanashvily:1992nr,Prugovecki:1995tj,Blagojevic:2013xpa} endeavours\textemdash seeking ``the" unified theory\textemdash have been striving to ``shoehorn" general relativity into the framework of quantum field theories (QFT).  Irrespective of the back-and-forth objections \cite{Ohanian:1995uu} and rebuttals \cite{Hehl:1997bz} on the torsion-based and Poincar\'{e} gauge geometrical approaches, what captures attention is the capability of these approaches to maintain \emph{Cosmic Censorship} \cite{Penrose:1999vj}, i.e., they can avoid nonphysical singularities resulted from black holes evaporation, and remove the ultraviolet divergences in QFT through treating fermions as spatially extended objects rather than perceiving them as ``point-like" particles \cite{Poplawski:2009su,Poplawski:2010kb,Poplawski:2011jz}, the mainstream QFT perspective on those particles. Part of this letter sheds some light on an alternative to this mainstream perspective as the approach we follow concurs with those geometrical approaches in viewing particles as dimensional objects.\\

In contrast, all non-geometrical attempts have shown drawbacks when it comes to deal with divergences and renormalization problems \cite{Eppley:1977fp,Shomer:2007vq,Albers:2008as}, especially when the mass of black hole $M_{BH}$ becomes closer to Planck mass $M_{P}$ at temperature $T_{BH}$ relating the mass of radiating black hole and its entropy. Such black hole is known as \textit{microscopic black hole}. If $M_{BH}<T_{BH}$ limit can be experimentally attained in high energy particle collisions, then studying the emitted particles from decay process of microscopic black holes would be very promising to divulge many secrets about how nature works on quantum gravity level \cite{Hsu:2002bd}. A black hole with mass of Sun would have $T_{BH}\sim 60\times 10^{-9}$ K at the event horizon, while a black hole with mass of Earth's moon would have the surface temperature $\sim 2.7$ K. So the radiation of astrophysical massive black hole is too minute to be detected in the Cosmic Microwave Background ($\sim 2.7$ K). In contrast, the end stage of primordial black holes comes with tidal and thermodynamical features that are more detectable as they are expected to be hotter, brighter and lighter ($M_{BH}\sim 10^{-8}$ kg) \cite{Hawking:1971ei, Page:1976wx}.\\

In light of the anticipated pattern of evaporation of tiny primordial black holes, it is also expected that ``man-made" microscopic black holes would evaporate after losing ``hair"\textemdash the presumably associated radiation fields\textemdash together with angular momentum and finally end up being with Schwarzschildian signatures and mass $\sim M_P$. This commutative scenario shows that Hawking temperature would have a divergent catastrophe when black holes are almost about to completely evaporate due to curvature singularities. However, this scenario overlooks the micro-structure quantum fluctuations of spacetime to taken into consideration. So it is suggested that it still can be maneuvered around with the help of noncommutative contrivances, including the proposal of extra dimensions \cite{Spallucci:2014kua}. Before we examine the significance of extra dimensions as a consequence of spacetime micro-structure \cite{Bleicher:2010qr}, it is worth noting some shortcomings of extra dimensions as a consequence of string theory alone without considering noncommutativity.\\

 The Arkani-Hamed, Dvali, Dimopoulos (ADD) model~\cite{ArkaniHamed:1998rs, ArkaniHamed:1998nn} predicts that a black hole with radius smaller than the size of the extra dimensions ($n>3$) can be placed in a (1, $n$)-dimensional isotropic spacetime, i.e., they are close to be microscopic black holes. Then higher-dimensional Schwarzschild solution appears in the scenario of black hole evaporation. So if high energy particle collisions would create microscopic black holes, they might evoke higher values of associated cross section in the presence of such large extra dimensions \cite{Bleicher:2007hw}. It is worth pointing out that ADD-based string theory models suggest quantum description for only extremal and super-extremal charged black hole models \cite{Strominger:1996sh}. However, it raises questions about the early discharge phase before reaching Schwarzschild geometry. Besides that, and still within the realm of string theory, there is no program that describes all evaporation phases of (super)-extremal black holes, specially the phase when the mass of black hole becomes $\sim M_P$ \cite{Nicolini:2008aj}.\\

Earlier before the introduction of \textit{large} extra dimensions in ADD model, it was proposed \cite{Madore:1989ma, Chamseddine:1992yx, Madore:1993fn} that the introduction of Noncommutative geometry (NC) in quantum gravity theories should imply extra dimensions. The proposal extends to substantially relate NC to the essential quantum fluctuations of gravitational field \cite{Madore:1995cg} by showing that classical gravity is indeed unique ``shadow" in the commutative limit of the noncommutative ``Fuzzy Spacetime'' \cite{Madore:1996bb, Madore:1996gr, Madore:1996sk, Madore:1997ta, Violette:1997ag}. Almost at the same time, Virtual Black Holes (VBH) were introduced to appear/disappear due to quantum fluctuations of spacetime too \cite{Hawking:1995ag, Wald:1998de, Crowell:2005ax} as consequence of relating uncertainty principle to Einstein equations of gravity such that VBH would \textit{gravitationally} resemble particle-antiparticle pairing in vacuum state of QFT \cite{Huggett:1998sz}. In light of this proposal, a VBH is to carry a mass $\sim M_P$ and to share features of Wheeler's quantum foam \cite{Wheeler:1955zz, Rickles:2018aoo}. So taking into account that uncertainty principle is a noncommutative relation that presumably forestalls measuring physical lengths more accurately than Planck length, and by considering all drawbacks of QFT in curved spacetime together with question marks on string theory predictions for extremal black hole, Nicolini black holes \cite{Nicolini:2005de, Nicolini:2005zi, Nicolini:2005vd, Rizzo:2006zb, Spallucci:2006zj, Ansoldi:2006vg, Spallucci:2009zz, Nicolini:2008aj, Casadio:2008qy, Arraut:2009an, Nicolini:2009gw, Gingrich:2010ed, Arraut:2010qx, Nicolini:2010nb} were introduced as a potential alternative to describe the end stage of primordial black holes with mass $\sim M_P$ in \emph{NC background}. Further details on noncommutative black holes and fuzzy geometry are in Ref. \cite{Nicolini:2008aj}. Also the rotating case \cite{Modesto:2010rv}, charged case \cite{Romero-Ayala:2015fba} and potential connection to primordial black holes \cite{Mann:2011mm} are considered. The case of Nicolini black hole with Schwarzschild geometry and VBH features is what we focus on through this letter.\\

One of the major issues with the many current approaches to quantum gravity research is the need for phenomenological features of a given quantum gravity model. It is the issue of having a set of observational/experimental constraints that allows eliminating some of the many other quantum gravity models.
Any quantum gravity phenomenology ought to be connected to the micro-structure of spacetime, such as spin foam models of canonical quantum gravity~\cite{Hawking:1979zw,Perez:2003vx,Garattini:2001yb,Baez:1997zt} or the compactified extra dimensions from brane world models and string theory~\cite{Randall:1999ee,Randall:2005xy,Antoniadis:1998ig}.
Major quantum gravity models phenomenologically result in
noncommutative geometry, where the conventional spacetime points are in a given
coordinate system~\cite{Seiberg:1999vs,Ardalan:1998ce,Aastrup:2012jj}. Furthermore, they form an algebra satisfying the Lie
bracket 
\be
\left[x^\mu, x^\nu\right] = i \theta g^{\mu \nu}
\ee
for~$\theta$ the noncommutativity parameter, with some matrix element~$g^{\mu\nu}$. The
implication for the above relation on point-like objects is to smear out such object into a Gaussian with a width~$\sqrt{\theta}$.\\
%
%

This \textit{natural} assumption about the micro-structure of the spacetime is based upon two main reasons.
The first one is to avoid the controversy of having undefined point-like particles, e.g.: electrons. This Boscovichian-like model results in characterizing those particles by an infinitely electromagnetic mass density. The only way to clear out such divergences is to use renormalization techniques, which essentially impose an effective cut-off scale for the quantum electrodynamics, which is a QFT, and hence avoiding indirectly the notion of point-like particles~\cite{Schwinger:1948iu}. It is also worth mentioning that due to the same reason, there is a new line of research has been launched \cite{Hooft:2016cpw} to describe black holes the same way physics describes particles with spatial volumes. Part of it refers indirectly to the assumed relationship that might be between VBH and NC. Bekenstein argued for similar argument \cite{Bekenstein:1997bt} although it targets different problem. Also torsion-based gauge theories of gravitation we mentioned earlier are endowed with noncommutative geometrical variety\textemdash but it is of a different kind as gravitational gauge theories are based on diffeomorphisms rather than Lie group structure of noncommutative coherent state formalism of spacetime\textemdash which what makes these theories see elementary particles with non-Boscovichian signature. This suggests that noncommutativity, in general, may be essential to the existence of elementary particles as spatially extended objects. The second reason is that the existence of a causal metric theory of gravity governed by the Einstein equations
%
%
%
along with localized spacetime events\textemdash that are determined by quantum radiation/matter interactions\textemdash would strongly recommend considering spacetime as a foam-like structure. This recommendation comes from the two known facts that spacetime obeys the uncertainty principle between position and momentum, and the Einstein equations imply the~\emph{ultra-relativistic} dispersion relation between energy and momentum~$ E\sim p$. Moreover, the existence of highly-localized energy would cause the spacetime structure to break-down beyond the Planck scale~\cite{Frohlich:1996zc}.\\

Another motivation for noncommutative geometry~\cite{PMIHES_1985__62__41_0} is the discovery of the area law for entropy~\cite{Bandyopadhyay:2003nu},
setting a bound on the maximum number of particle/events in a given region
of spacetime bounded by an area~$A$
\be
N\lesssim \frac{A}{\ell_p^2}~.
\ee
Hence from the previous discussion, we would expect that the spacetime at the
micro-scale to consist of a sea of VBH~\cite{Faizal:2006cm,Hawking:1995ag}. However,
in all the models studying VBH, the noncommutative structure of spacetime has not been taken into an account, a priori, although the motivation is the same for both phenomena. VBH could have a measurable effect in particle physics, permitting events/decays that are forbidden within the realm of the standard model. The most important decay that could be caused by VBH is the proton decay~\cite{Adams:2000za}. Noncommutative
spacetime models also predict phenomenological aspects on particle physics, but they seem to be rather ill-defined or even unjustified.
A widely known prediction of noncommutative spacetime geometry is the
mass of the Higgs particles~$m_H$. It is predicted to be about~$\sqrt{2}m_t \sim246$ GeV, where $m_t$ is the mass of the top quark~\cite{Connes:1990qp}. This
has a considerable error compared to the measured mass~$m_H=125$ GeV but remains in the same order of magnitude.  Given the time these calculations
were made, the predictions from noncommutative geometry seemed within the
experimental range.
\begin{figure}[h!]
	\centering 
	\label{timeline}
	\includegraphics[width=\linewidth]{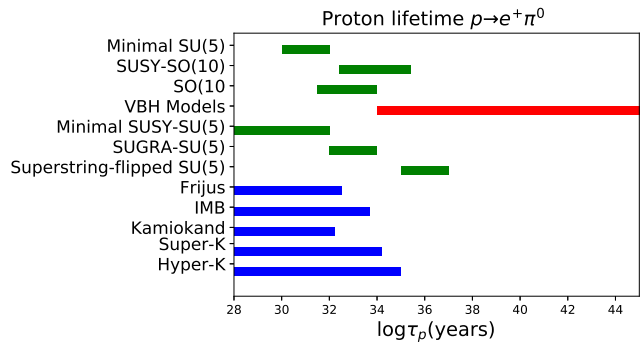}
	\caption{Theoretical predictions for the proton lifetime $\tau_p$ in the most prominent models Vs. the experimental searches for the proton decay. Observations show that most of these models has been ruled out, leaving a tight window for $ SO(10)$ and flipped $SU(5)$ models. In addition to the VBH models that can be tuned for a large range.~\cite{Adams:2000za,Nishino:2009aa,Gajewski:1981kv,Dimopoulos:1981dw,Sakai:1981pk,Bajc:2002bv,Frampton:1990hz,Langacker:1980js}  }
\end{figure}

Considering the experimental/observational bound of the lifetime of the
proton $ \tau_p > 10^{34}$ years~\cite{Nishino:2009aa}, then many models such as grand unification theories (GUT)\footnote{In GUT, magnetic monopoles are interesting example of processes catalyze proton decay, particularly the monopoles of SU(N) with the Rubakov-Callan effect \cite{Rubakov:1981rg, Callan:1982ac}. Those monopoles should be differentiated from those of SO(N) dual gravity \cite{Danehkar:2019qmw, Curtright:2019yur, Curtright:2019wxg, Alshal:2019hpk}, even if SU(5) of Georgi-Glashaw model can be embedded within SO(10) \cite{Dawson:1982sc}.}, supersymmetric models (MSSM's in particular)~\cite{Georgi:1974sy,Dimopoulos:1981dw,Sakai:1981pk,Bajc:2002bv} or sphaleron model~\cite{Arnold:1987mh} would be eliminated from the consideration. However, such consideration would leave a
tight window for other GUT models, in particular, like those involving strings and branes~\cite{Nath:2006ut}, variations of SO(10) group~\cite{Baez:2009dj}, or by leptoquarks models~\cite{Dorsner:2012nq} which shows an increasing interest, due to recent findings related to the anomalies in $B$-meson decays at the LHCb experiment~\cite{Aaij:2014ora}. Moreover, the proton could decay via VBH, and the lifetime of this decay can be estimated from the relation in $D$ dimensions~\cite{Alsaleh:2017ttv,Adams:2000za}
\be
\tau_p \sim M_{proton}^{-1} \left(\frac{M_{qg}}{M_{proton}}\right)^D,
\label{lifetime}
\ee
where, for the VBH mass~$ M_{qg} = \sqrt{1 / 8 \pi G} = M_p$ the Planck mass and $D=4$, the proton lifetime ~$\tau$ is $\sim 10^{45}$ years. Not to mention that the proton decay process, if it exists, is
a very rare event, therefore, it also gives branching ratio~$\mathcal{B}$ between the generic GUT decay channel and QG one of order $\mathcal{B} = 10^{19}$ which is extremely small. However, for different models of quantum gravity and extra dimensions, the VBH channel would have a significant contribution to the proton decay. In fact, for phenomenological quantum gravity models such as generalized uncertainty principle~(GUP)\cite{Amati:1988tn, Garay:1994en, Kempf:1994su, Adler:2001vs, Ali:2009zq, Vagenas:2018zoz, Vagenas:2018pez, Vagenas:2019wzd, Vagenas:2019rai} the VBH decay channel could have comparable effects to GUP/SUSY or other models for reasonable deformation parameter $ \beta$~\cite{Alsaleh:2017ttv}. Since the experimental researches have excluded many non-quantum gravity models, the possibility that proton decay being a signature of quantum gravity is increasing, see FIG. 1.~\ref{timeline}

\section*{Gaussian distribution for the mass of virtual black holes in noncommutative geometrical background}
 It would be interesting to investigate the hypothesis of noncommutative spacetime as a phenomenological quantum gravity model on the proton decay via noncommutative VBH~(NCVBH) and to examine the experimental limits on the noncommutativity parameter~$\theta$. The mass density distribution for a droplet of matter/equivalent of energy in $D$ space time dimension is given by \cite{Nicolini:2005vd,Tejeiro:2010gu}.
\be
\rho(r) = \frac{M}{\left( 4 \pi \theta \right) ^{D-1/2}} \, e^{-\frac{r^2}{4 \theta }}.
\ee

  Assuming a Gaussian mass density distribution with width $ \theta$, we begin studying the geometric properties of noncommutative VBH by computing the Einstein field equations using the metric of a microscopic black hole \cite{Nicolini:2005vd}.
\be
-ds^2 = & \left( 1-\frac{4M}{r \sqrt{\pi}} \gamma(3/2,r^2/4\theta)\right) dt^2\\
& -\left( 1-\frac{4M}{r \sqrt{\pi}}\, \gamma(3/2,r^2/4\theta)\right)^{-1}dr^2-r^2 d\Omega^2_2~,
\label{NCVBH}
\ee
where $ M$ is the black hole mass, which is an unknown parameter in this case. And~$\gamma(3/2,4\theta)$ is the incomplete gamma function.
\be
\gamma(3/2,r^2/4\theta) = \int_{0}^{r^2/4\theta}t^{1/2}e^{-t} dt.
\ee

Following the analysis in~\cite{Arraut:2009an,Nicolini:2005vd}, we find the spacetime metric for $4$ dimensions, and we can generalize this analysis for arbitrary $D$ dimensions.

\begin{figure}[h!]
	\centering 
	\label{plt_rand}
	\includegraphics[width=0.5\textwidth]{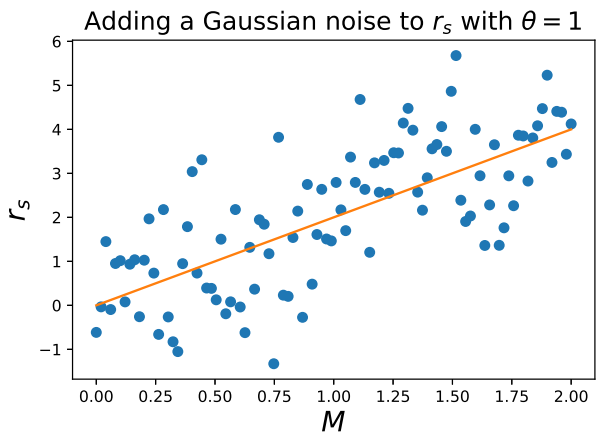}
	\caption{The standard gravitational radius~$ r_s=2M$ with added Gaussian noise of standard deviation $ \sqrt{\theta}=1$. The linear relationship is described by $r_s=2M+\delta r$, where $\delta r$ represents the standard error generated by normally distributed random error with a mean $ = 0$ and 100 random trials. Notice that the negative radius region is nonphysical, hence~$M$ must be at least~$ M> \sqrt{\theta}$.}
\end{figure}

 This metric gives a noncommutative gravitational radius $ r_\Delta$ that would be of concern when we examine the proton lifetime. Nevertheless, the mass of (virtual) black hole needs further study in order to identify it. \\ We want the above metric~\eqref{NCVBH} to become the conventional Schwarzschild metric when $ \theta \to 0$ with a Planck mass~$M_p$ being its mass. This ansatz is attainable as the limit $ \gamma(a,z) \to \Gamma(a)$ as $ z\to \infty$~\cite{Abramowitz:1974:HMF:1098650}. So $M$ is, indeed, Planck mass.\\ Also we identify the effective gravitational radius $ r_s$ (or equivalently the effective quantum gravity mass~$M_{qg}$) as the solution to the equation
 \be
h(r)|_{r_s}:=\frac{4M}{r_s \sqrt{\pi}}\, \gamma(3/2,r_s^2/4\theta)-1=0.
\label{hor}
 \ee
In this present form, we could not analytically solve this equation for $r_s$. Therefore, we expand the incomplete gamma function at the classical spacetime limit~$ \theta \to 0$ and take the leading and sub-leading terms. Then the incomplete gamma function is expanded as~\cite{Abramowitz:1974:HMF:1098650}
\be
\gamma(3/2,r^2/4\theta) \simeq \Gamma(3/2) -\frac{2}{\sqrt{\theta}}r e^{-r^2/4\theta} + \mathcal{O}(\theta ^{1/2}),
\ee
which is rather an expected result. The sub-leading term is Gaussian, superimposing Gaussian distribution \textit{noise} around the standard gravitational radius~$ r_s = 2M$, see the plot in FIG. 2.~\ref{plt_rand}

\noindent Therefore, the horizon equation can be written as up to sub-leading order as
\be
h(r)|_{r_s}=r_s - r_s\,\mathcal N (r; r_s,\sqrt{\theta}),
\ee
with~$\mathcal N (r; r_s,\sqrt{\theta})$ being the normal distribution. Then from \cite{Nicolini:2005vd} with $r_s=2M$ and standard deviation $ \sigma = \sqrt{\theta}$, we can directly define the minimal effective gravitational mass to be $ M_{gq}\sim \sqrt{\theta}$ since the noncommutative black hole could not be defined with radius less that $\sqrt{\theta}$, i.e., at short distances we consider the quantum geometry effects made by spacetime fuzziness where $r_s \sim \sqrt{\theta}$. This suggests a consistent picture to present the basic phenomenology of quantum gravity, particularly the description of VBH, without setting an artificial bounds on the gravitational mass/radius.\\ Moreover, this result can be realized within the stochastic interpretation of quantization~\cite{Damgaard:1988nq,Bandyopadhyay:2003nu} which assumes that the gravitational degree of freedom is the black hole's gravitational radius $r_s$ with the rate of its change as the conjugate momentum does. Therefore, we add to them a stochastic extension with the parameter $ \sqrt{\theta} $ such that
\be
r_s +i \sqrt{\theta}\hat{Q},
\ee
which is similar to what was obtained in the formal scholastic quantization of black holes by Moffat~\cite{Moffat:1996fu,Moffat:2014eua}.

\section*{Numerical analysis in $D$ dimensions}
\begin{figure}[h!]
	\centering 
	\label{plt_gamma}
	\includegraphics[width=0.9\linewidth]{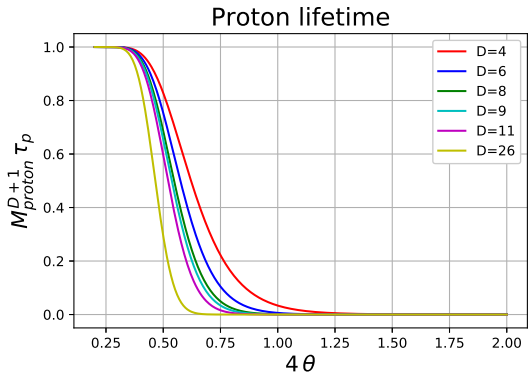}
	\caption{The proton lifetime as a function of the noncommutativity parameter~$\theta$ in Planck units for $4,6,8,9,11$ and $ 26$ spacetime dimensions $D$. }
\end{figure}

\begin{figure}[t!]
	\centering 
	\label{plt_dim}
	\includegraphics[width=0.9\linewidth]{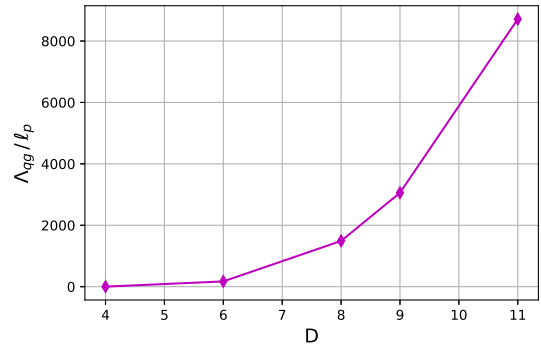}
	\caption{Visualization of the relation between the spacetime dimensions $D$ and the minimum noncommutativity scale $ \Lambda_{qg}/\ell_p$}
\end{figure}

 From this analysis we could rewrite $\theta$ in terms of the effective scale of quantum gravity, that is $ \Lambda_{QG}= \frac{1}{\sqrt{\theta}}$. And since the virtual black hole mass is bounded by the noncommutativity parameter, we can recover the result of the virtual black hole mass being corresponded to the effective scale of quantum gravity~$ \Lambda_{qg}\sim \frac{1}{M_{VBH}}$. Therefore, if observations were made with careful analysis, a crucial observation of black hole decay would reveal the micro-structure of spacetime. The experimental and observational bound of the minimal mass of black holes can be found in~\cite{Abazov:2008kp,Gingrich:2006hm,Aaltonen:2008hh,Khachatryan:2010wx} where the mass is bounded to be $ > 4.5$ TeV. According to our analysis, that corresponds to a quantum gravity scale bound of $\sim4.39 \times 10^{-20}$ m, which is clearly much larger from what we expect, as this scale is comparable to the electroweak scale that does not show spacetime anti-commutativity at it. Other models excluded the possibility of detecting microscopic black holes at the LHC even at run II with $ \sqrt{s}= 14 $ TeV due to phenomenology of quantum gravity, such as modified dispersion relation by rainbow functions, or existence of maximum momentum by a generalization of Heisenberg algebra GUP~\cite{Ali:2012mt,Cavaglia:2003qk}. These bounds were set using particle collisions. However, the proton lifetime could set a much better bound if the relation~\eqref{lifetime} is used and the $M_{qg}$ with the quantum gravity scale $ \Lambda_{QG}^{-1}$ is substituted. This leads to the order of unity estimation. Using the relation~\eqref{hor} in the proton lifetime formula, we can numerically find the bound on the noncommutativity parameter~$\theta$. Alternatively, the bound on the noncommutativity scale~(quantum gravity scale)~$\Lambda_{qg}=1/\sqrt{\theta}$ can be estimated from the experimental bound of the proton lifetime $ > 10^{34}$ years. The relation is between $\theta$ and the mass of the proton to the $D+1$ power, multiplied with $\tau_p$ is shown in FIG. 3.~\ref{plt_gamma} Numerical computations results are summarized in table. I.~\ref{result} and visualized in FIG. 4.~\ref{plt_dim}

\vspace{1cm}
\begin{table}[h!]
\scalebox{1.5}{
	\label{result}
	\begin{tabular}{|c|c|}
		\hline
	 $D$ &$\Lambda_{qg}/\ell_p$ \\
		\hline
		4 & $2.269$ \\
		6 & $171.940$ \\
		8 & $1.489 \times 10^{3}$ \\
		9 & $3.059 \times 10^{3}$\\
		11 &$ 8.714\times 10^{3}$ \\
		26 & $1.320 \times 10^{5}$ \\
		\hline
	\end{tabular}
	}
\caption{The bound on the quantum gravity noncommutativity scale $ \Lambda_{qg}$ in Planck length units, from the observational bound on the proton lifetime. For different spacetime dimensions}
\end{table}

\section*{Conclusions}
We investigated the proton lifetime and how experimental results showed the non-validity of many quantum gravity models. We suggested perceiving the decay of proton as the thermal evaporation of virtual black holes within the context of noncommutative geometry. We used the lower incomplete gamma function to relate the Gaussian distribution of the mass density to the mass of Schwarzschild-like virtual black holes, and we calculated both corresponding mass and gravitational radius of the horizon of such black hole in terms of the noncommutativity parameter $\theta$. This introduced an experimental verifiable way to check the validity of seeing the micro-structure of the spacetime in the context of the noncommutative geometry. Finally, we numerically analyzed the process of decay in different $D$ dimensions, and we showed the possible bounds on the noncommutativity parameter~$\theta$. The study can be extended for investigating the implications of noncommutativity geometry in cosmology to be compared with the recent Planck data. We hope to report on this in the future.

\section*{Acknowledgment}

Authors would like to thank the anonymous reviewers of the manuscript for their constructive suggestions to amend the presentation of the letter.\\
 A.A and S.A. were supported by a grant from the “Research Center of the Female Scientific and Medical Colleges”, Deanship of Scientific Research, King Saud University.
%


\bibliography{ref}

\begin{thebibliography}{100}

\bibitem{Hawking:1974sw}
S.~W. Hawking.
\newblock {Particle Creation by Black Holes}.
\newblock {\em Commun. Math. Phys.}, 43:199--220, 1975.
\newblock [,167(1975)].

\bibitem{Birrell:1982ix}
N.~D. Birrell and P.~C.~W. Davies.
\newblock {\em {Quantum Fields in Curved Space}}.
\newblock Cambridge Monographs on Mathematical Physics. Cambridge Univ. Press,
  Cambridge, UK, 1984.

\bibitem{DeWitt:2003pm}
Bryce~S. DeWitt.
\newblock {The global approach to quantum field theory. Vol. 1, 2}.
\newblock {\em Int. Ser. Monogr. Phys.}, 114:1--1042, 2003.

\bibitem{Fulling:1989nb}
S.~A. Fulling.
\newblock {Aspects of Quantum Field Theory in Curved Space-time}.
\newblock {\em London Math. Soc. Student Texts}, 17:1--315, 1989.

\bibitem{Wald:1995yp}
Robert~M. Wald.
\newblock {\em {Quantum Field Theory in Curved Space-Time and Black Hole
  Thermodynamics}}.
\newblock Chicago Lectures in Physics. University of Chicago Press, Chicago,
  IL, 1995.

\bibitem{Ford:1997hb}
L.~H. Ford.
\newblock {Quantum field theory in curved space-time}.
\newblock In {\em {Particles and fields. Proceedings, 9th Jorge Andre Swieca
  Summer School, Campos do Jordao, Brazil, February 16-28, 1997}}, pages
  345--388, 1997.

\bibitem{Bar:2009zzb}
C.~Bär and K.~Fredenhagen.
\newblock {Quantum field theory on curved spacetimes}.
\newblock {\em Lect. Notes Phys.}, 786:1--155, 2009.

\bibitem{Parker:2009uva}
Leonard~E. Parker and D.~Toms.
\newblock {\em {Quantum Field Theory in Curved Spacetime}}.
\newblock Cambridge Monographs on Mathematical Physics. Cambridge University
  Press, 2009.

\bibitem{Donoghue:2017pgk}
John~F. Donoghue, Mikhail~M. Ivanov, and Andrey Shkerin.
\newblock {EPFL Lectures on General Relativity as a Quantum Field Theory}.
\newblock 2017.

\bibitem{Sardanashvily:1992nr}
G.~A. Sardanashvily and O.~Zakharov.
\newblock {\em {Gauge gravitation theory}}.
\newblock 1992.

\bibitem{Prugovecki:1995tj}
E.~Prugovecki.
\newblock {\em {Principles of quantum general relativity}}.
\newblock 1995.

\bibitem{Blagojevic:2013xpa}
Milutin Blagojević and Friedrich~W. Hehl, editors.
\newblock {\em {Gauge Theories of Gravitation}}.
\newblock World Scientific, Singapore, 2013.

\bibitem{Ohanian:1995uu}
H.~Ohanian and R.~Ruffini.
\newblock {\em {Gravitation and space-time},}.
\newblock 1995, p.228-229.

\bibitem{Hehl:1997bz}
Friedrich~W. Hehl.
\newblock {Alternative gravitational theories in four-dimensions}.
\newblock In {\em {Recent developments in theoretical and experimental general
  relativity, gravitation, and relativistic field theories. Proceedings, 8th
  Marcel Grossmann meeting, MG8, Jerusalem, Israel, June 22-27, 1997. Pts. A,
  B}}, pages 423--432, 1997.

\bibitem{Penrose:1999vj}
Roger Penrose.
\newblock {The question of cosmic censorship}.
\newblock {\em J. Astrophys. Astron.}, 20:233--248, 1999.

\bibitem{Poplawski:2009su}
Nikodem~J. Poplawski.
\newblock {Nonsingular Dirac particles in spacetime with torsion}.
\newblock {\em Phys. Lett.}, B690:73--77, 2010.
\newblock [Erratum: Phys. Lett.B727,575(2013)].

\bibitem{Poplawski:2010kb}
Nikodem~J. Popławski.
\newblock {Cosmology with torsion: An alternative to cosmic inflation}.
\newblock {\em Phys. Lett.}, B694:181--185, 2010.
\newblock [Erratum: Phys. Lett.B701,672(2011)].

\bibitem{Poplawski:2011jz}
Nikodem~J. Poplawski.
\newblock {Nonsingular, big-bounce cosmology from spinor-torsion coupling}.
\newblock {\em Phys. Rev.}, D85:107502, 2012.

\bibitem{Eppley:1977fp}
K.~{Eppley} and E.~{Hannah}.
\newblock {The necessity of quantizing the gravitational field}.
\newblock {\em Foundations of Physics}, 7:51--68, February 1977.

\bibitem{Shomer:2007vq}
Assaf Shomer.
\newblock {A Pedagogical explanation for the non-renormalizability of gravity}.
\newblock 2007.

\bibitem{Albers:2008as}
Mark Albers, Claus Kiefer, and Marcel Reginatto.
\newblock {Measurement Analysis and Quantum Gravity}.
\newblock {\em Phys. Rev.}, D78:064051, 2008.

\bibitem{Hsu:2002bd}
Stephen D.~H. Hsu.
\newblock {Quantum production of black holes}.
\newblock {\em Phys. Lett.}, B555:92--98, 2003.

\bibitem{Hawking:1971ei}
Stephen Hawking.
\newblock {Gravitationally collapsed objects of very low mass}.
\newblock {\em Mon. Not. Roy. Astron. Soc.}, 152:75, 1971.

\bibitem{Page:1976wx}
Don~N. Page and S.~W. Hawking.
\newblock {Gamma rays from primordial black holes}.
\newblock {\em Astrophys. J.}, 206:1--7, 1976.

\bibitem{Spallucci:2014kua}
Euro Spallucci and Anais Smailagic.
\newblock {Semi-classical approach to quantum black holes}.
\newblock In {\em {"Advances in black holes research'' p.1-26, Ed.: A. Barton,
  Nova Science Publisher, Inc. (2015), ISBN: 978-1-63463-168-6}}, 2014.

\bibitem{Bleicher:2010qr}
Marcus Bleicher and Piero Nicolini.
\newblock {Large Extra Dimensions and Small Black Holes at the LHC}.
\newblock {\em J. Phys. Conf. Ser.}, 237:012008, 2010.

\bibitem{ArkaniHamed:1998rs}
Nima Arkani-Hamed, Savas Dimopoulos, and G.~R. Dvali.
\newblock {The Hierarchy problem and new dimensions at a millimeter}.
\newblock {\em Phys. Lett.}, B429:263--272, 1998.

\bibitem{ArkaniHamed:1998nn}
Nima Arkani-Hamed, Savas Dimopoulos, and G.~R. Dvali.
\newblock {Phenomenology, astrophysics and cosmology of theories with
  submillimeter dimensions and TeV scale quantum gravity}.
\newblock {\em Phys. Rev.}, D59:086004, 1999.

\bibitem{Bleicher:2007hw}
Marcus Bleicher.
\newblock {How to Create Black Holes on Earth?}
\newblock {\em Eur. J. Phys.}, 28:509--516, 2007.

\bibitem{Strominger:1996sh}
A.~Strominger and C.~Vafa.
\newblock {Microscopic origin of the Bekenstein-Hawking entropy}.
\newblock {\em Phys. Lett.}, B379:99--104, 1996.

\bibitem{Nicolini:2008aj}
Piero Nicolini.
\newblock {Noncommutative Black Holes, The Final Appeal To Quantum Gravity: A
  Review}.
\newblock {\em Int. J. Mod. Phys.}, A24:1229--1308, 2009.

\bibitem{Madore:1989ma}
J.~Madore.
\newblock {On a Modification of {Kaluza-Klein} Theory}.
\newblock {\em Phys. Rev.}, D41:3709, 1990.

\bibitem{Chamseddine:1992yx}
Ali~H. Chamseddine, Giovanni Felder, and J.~Frohlich.
\newblock {Gravity in noncommutative geometry}.
\newblock {\em Commun. Math. Phys.}, 155:205--218, 1993.

\bibitem{Madore:1993fn}
J.~Madore and J.~Mourad.
\newblock {Algebraic Kaluza-Klein cosmology}.
\newblock {\em Class. Quant. Grav.}, 10:2157--2170, 1993.

\bibitem{Madore:1995cg}
J.~Madore and J.~Mourad.
\newblock {On the origin of Kaluza-Klein structure}.
\newblock {\em Phys. Lett.}, B359:43--48, 1995.

\bibitem{Madore:1996bb}
J.~Madore and J.~Mourad.
\newblock {Quantum space-time and classical gravity}.
\newblock {\em J. Math. Phys.}, 39:423--442, 1998.

\bibitem{Madore:1996gr}
J.~Madore.
\newblock {Fuzzy space-time}.
\newblock {\em Can. J. Phys.}, 75:385, 1997.

\bibitem{Madore:1996sk}
J.~Madore.
\newblock {Classical gravity on fuzzy space-time}.
\newblock {\em Nucl. Phys. Proc. Suppl.}, 56B(3):183--190, 1997.

\bibitem{Madore:1997ta}
J.~Madore.
\newblock {Gravity on fuzzy space-time}.
\newblock In {\em {International Workshop on Mathematical Physics - Today,
  Priority Technologies - for Tomorrow Kiev, Ukraine, May 12-17, 1997}}, 1997.

\bibitem{Violette:1997ag}
John Madore~J. Dubois-Violette, M. and R.~Kerner.
\newblock {Shadow of noncommutativity }, 1997.

\bibitem{Hawking:1995ag}
S.~W. Hawking.
\newblock {Virtual black holes}.
\newblock {\em Phys. Rev.}, D53:3099--3107, 1996.

\bibitem{Wald:1998de}
Robert~M. Wald, editor.
\newblock {\em {Black holes and relativistic stars}}.
\newblock 1998.

\bibitem{Crowell:2005ax}
L.~B. Crowell.
\newblock {\em {Quantum fluctuations of spacetime}}.
\newblock 2005.

\bibitem{Huggett:1998sz}
S.~A. Huggett, L.~J. Mason, K.~P. Tod, S.~T. Tsou, and N.~M.~J. Woodhouse,
  editors.
\newblock {\em {The geometric universe: Science, geometry, and the work of
  Roger Penrose. Proceedings, Symposium, Geometric Issues in the Foundations of
  Science, Oxford, UK, June 25-29, 1996}}, 1998.

\bibitem{Wheeler:1955zz}
J.~A. Wheeler.
\newblock {Geons}.
\newblock {\em Phys. Rev.}, 97:511--536, 1955.

\bibitem{Rickles:2018aoo}
Dean Rickles.
\newblock {Geon Wheeler: from nuclear to spacetime physicist}.
\newblock {\em Eur. Phys. J.}, H43(3):243--265, 2018.

\bibitem{Nicolini:2005de}
Piero Nicolini, Anais Smailagic, and Euro Spallucci.
\newblock {The Fate of radiating black holes in noncommutative geometry}.
\newblock 2005.
\newblock [ESA Spec. Publ.637,11.1(2006)].

\bibitem{Nicolini:2005zi}
Piero Nicolini.
\newblock {A Model of radiating black hole in noncommutative geometry}.
\newblock {\em J. Phys.}, A38:L631--L638, 2005.

\bibitem{Nicolini:2005vd}
Piero Nicolini, Anais Smailagic, and Euro Spallucci.
\newblock {Noncommutative geometry inspired Schwarzschild black hole}.
\newblock {\em Phys. Lett.}, B632:547--551, 2006.

\bibitem{Rizzo:2006zb}
Thomas~G. Rizzo.
\newblock {Noncommutative Inspired Black Holes in Extra Dimensions}.
\newblock {\em JHEP}, 09:021, 2006.

\bibitem{Spallucci:2006zj}
Euro Spallucci, Anais Smailagic, and Piero Nicolini.
\newblock {Trace Anomaly in Quantum Spacetime Manifold}.
\newblock {\em Phys. Rev.}, D73:084004, 2006.

\bibitem{Ansoldi:2006vg}
Stefano Ansoldi, Piero Nicolini, Anais Smailagic, and Euro Spallucci.
\newblock {Noncommutative geometry inspired charged black holes}.
\newblock {\em Phys. Lett.}, B645:261--266, 2007.

\bibitem{Spallucci:2009zz}
Euro Spallucci, Anais Smailagic, and Piero Nicolini.
\newblock {Non-commutative geometry inspired higher-dimensional charged black
  holes}.
\newblock {\em Phys. Lett.}, B670:449--454, 2009.

\bibitem{Casadio:2008qy}
Roberto Casadio and Piero Nicolini.
\newblock {The Decay-time of non-commutative micro-black holes}.
\newblock {\em JHEP}, 11:072, 2008.

\bibitem{Arraut:2009an}
I.~Arraut, Davide Batic, and Marek Nowakowski.
\newblock {A Non commutative model for a mini black hole}.
\newblock {\em Class. Quant. Grav.}, 26:245006, 2009.

\bibitem{Nicolini:2009gw}
Piero Nicolini and Euro Spallucci.
\newblock {Noncommutative geometry inspired wormholes and dirty black holes}.
\newblock {\em Class. Quant. Grav.}, 27:015010, 2010.

\bibitem{Gingrich:2010ed}
Douglas~M. Gingrich.
\newblock {Noncommutative geometry inspired black holes in higher dimensions at
  the LHC}.
\newblock {\em JHEP}, 05:022, 2010.

\bibitem{Arraut:2010qx}
I.~Arraut, D.~Batic, and M.~Nowakowski.
\newblock {Maximal Extension of the Schwarzschild Spacetime Inspired by
  Noncommutative Geometry}.
\newblock {\em J. Math. Phys.}, 51:022503, 2010.

\bibitem{Nicolini:2010nb}
Piero Nicolini.
\newblock {Entropic force, noncommutative gravity and un-gravity}.
\newblock {\em Phys. Rev.}, D82:044030, 2010.

\bibitem{Modesto:2010rv}
Leonardo Modesto and Piero Nicolini.
\newblock {Charged rotating noncommutative black holes}.
\newblock {\em Phys. Rev.}, D82:104035, 2010.

\bibitem{Romero-Ayala:2015fba}
Carlos~A. Soto-Campos and Susana Valdez-Alvarado.
\newblock {Noncommutative Reissner-Nordstr{\o}m Black hole}.
\newblock 2015.

\bibitem{Mann:2011mm}
Robert~B. Mann and Piero Nicolini.
\newblock {Cosmological production of noncommutative black holes}.
\newblock {\em Phys. Rev.}, D84:064014, 2011.

\bibitem{Hawking:1979zw}
S.~W. Hawking.
\newblock {Space-Time Foam}.
\newblock {\em Nucl. Phys.}, B144:349--362, 1978.

\bibitem{Perez:2003vx}
Alejandro Perez.
\newblock {Spin foam models for quantum gravity}.
\newblock {\em Class. Quant. Grav.}, 20:R43, 2003.

\bibitem{Garattini:2001yb}
Remo Garattini.
\newblock {What Casimir energy can suggest about space-time foam?}
\newblock {\em Int. J. Mod. Phys.}, A17:829--832, 2002.

\bibitem{Baez:1997zt}
John~C. Baez.
\newblock {Spin foam models}.
\newblock {\em Class. Quant. Grav.}, 15:1827--1858, 1998.

\bibitem{Randall:1999ee}
Lisa Randall and Raman Sundrum.
\newblock {A Large mass hierarchy from a small extra dimension}.
\newblock {\em Phys. Rev. Lett.}, 83:3370--3373, 1999.

\bibitem{Randall:2005xy}
L.~Randall.
\newblock {\em {Warped passages: Unraveling the mysteries of the universe's
  hidden dimensions}}.
\newblock 2005.

\bibitem{Antoniadis:1998ig}
Ignatios Antoniadis, Nima Arkani-Hamed, Savas Dimopoulos, and G.~R. Dvali.
\newblock {New dimensions at a millimeter to a Fermi and superstrings at a
  TeV}.
\newblock {\em Phys. Lett.}, B436:257--263, 1998.

\bibitem{Seiberg:1999vs}
Nathan Seiberg and Edward Witten.
\newblock {String theory and noncommutative geometry}.
\newblock {\em JHEP}, 09:032, 1999.

\bibitem{Ardalan:1998ce}
F.~Ardalan, H.~Arfaei, and M.~M. Sheikh-Jabbari.
\newblock {Noncommutative geometry from strings and branes}.
\newblock {\em JHEP}, 02:016, 1999.

\bibitem{Aastrup:2012jj}
Johannes Aastrup and Jesper~Moller Grimstrup.
\newblock {Intersecting Quantum Gravity with Noncommutative Geometry: A
  Review}.
\newblock {\em SIGMA}, 8:018, 2012.

\bibitem{Schwinger:1948iu}
Julian~S. Schwinger.
\newblock {On Quantum electrodynamics and the magnetic moment of the electron}.
\newblock {\em Phys. Rev.}, 73:416--417, 1948.

\bibitem{Hooft:2016cpw}
Gerard 't~Hooft.
\newblock {The Quantum Black Hole as a Hydrogen Atom: Microstates Without
  Strings Attached}.
\newblock 2016.

\bibitem{Bekenstein:1997bt}
Jacob~D. Bekenstein.
\newblock {Quantum black holes as atoms}.
\newblock In {\em {Recent developments in theoretical and experimental general
  relativity, gravitation, and relativistic field theories. Proceedings, 8th
  Marcel Grossmann meeting, MG8, Jerusalem, Israel, June 22-27, 1997. Pts. A,
  B}}, pages 92--111, 1997.

\bibitem{Frohlich:1996zc}
J.~Frohlich, O.~Grandjean, and A.~Recknagel.
\newblock {Supersymmetric quantum theory and (noncommutative) differential
  geometry}.
\newblock {\em Commun. Math. Phys.}, 193:527--594, 1998.

\bibitem{PMIHES_1985__62__41_0}
Alain Connes.
\newblock Non-commutative differential geometry.
\newblock {\em Publications Math\'ematiques de l'IH\'ES}, 62:41--144, 1985.

\bibitem{Bandyopadhyay:2003nu}
P.~Bandyopadhyay.
\newblock {\em {Geometry, topology and quantum field theory}}, volume 130.
\newblock 2003.

\bibitem{Faizal:2006cm}
Mir Faizal.
\newblock {Some aspects of virtual black holes}.
\newblock {\em J. Exp. Theor. Phys.}, 114:400--405, 2012.

\bibitem{Adams:2000za}
Fred~C. Adams, Gordon~L. Kane, Manasse Mbonye, and Malcolm~J. Perry.
\newblock {Proton decay, black holes, and large extra dimensions}.
\newblock {\em Int. J. Mod. Phys.}, A16:2399--2410, 2001.

\bibitem{Connes:1990qp}
Alain Connes and John Lott.
\newblock {Particle Models and Noncommutative Geometry (Expanded Version)}.
\newblock {\em Nucl. Phys. Proc. Suppl.}, 18B:29--47, 1991.

\bibitem{Nishino:2009aa}
H.~Nishino et~al.
\newblock {Search for Proton Decay via {$p \to e^{+} \pi^{0}$} and {$p \to
  \mu^{+} \pi^{0}$} in a Large Water Cherenkov Detector}.
\newblock {\em Phys. Rev. Lett.}, 102:141801, 2009.

\bibitem{Gajewski:1981kv}
W.~Gajewski et~al.
\newblock {The IMB proton decay experiment.}
\newblock In {\em {NEUTRINO '81: Proceedings of the 9th International
  Conference on Neutrino Physics and Astrophysics Wailea, Hawaii July 1-8, 1981
  (Vol. 1)}}, pages 205--214, 1981.

\bibitem{Dimopoulos:1981dw}
Savas Dimopoulos, Stuart Raby, and Frank Wilczek.
\newblock {Proton Decay in Supersymmetric Models}.
\newblock {\em Phys. Lett.}, 112B:133, 1982.

\bibitem{Sakai:1981pk}
N.~Sakai and Tsutomu Yanagida.
\newblock {Proton Decay in a Class of Supersymmetric Grand Unified Models}.
\newblock {\em Nucl. Phys.}, B197:533, 1982.

\bibitem{Bajc:2002bv}
Borut Bajc, Pavel Fileviez~Perez, and Goran Senjanovic.
\newblock {Proton decay in minimal supersymmetric SU(5)}.
\newblock {\em Phys. Rev.}, D66:075005, 2002.

\bibitem{Frampton:1990hz}
Paul~H. Frampton and Thomas~W. Kephart.
\newblock {Higgs sector and proton decay in SU(15) grand unification}.
\newblock {\em Phys. Rev.}, D42:3892--3894, 1990.

\bibitem{Langacker:1980js}
Paul Langacker.
\newblock {Grand Unified Theories and Proton Decay}.
\newblock {\em Phys. Rept.}, 72:185, 1981.

\bibitem{Rubakov:1981rg}
V.~A. Rubakov.
\newblock {Superheavy Magnetic Monopoles and Proton Decay}.
\newblock {\em JETP Lett.}, 33:644--646, 1981.
\newblock [Pisma Zh. Eksp. Teor. Fiz.33,658(1981)].

\bibitem{Callan:1982ac}
Curtis~G. Callan, Jr.
\newblock {Monopole Catalysis of Baryon Decay}.
\newblock {\em Nucl. Phys.}, B212:391--400, 1983.

\bibitem{Danehkar:2019qmw}
Ashkbiz Danehkar.
\newblock {Electric-magnetic duality in gravity and higher-spin fields}.
\newblock {\em Front.in Phys.}, 6:146, 2019.

\bibitem{Curtright:2019yur}
Thomas~L. Curtright.
\newblock {Massive dual spinless fields revisited}.
\newblock {\em Nucl. Phys. B.}, 948:114784, 2019.

\bibitem{Curtright:2019wxg}
T.~L. Curtright and H.~Alshal.
\newblock {Massive dual spin 2 revisited}.
\newblock {\em Nucl. Phys. B.}, 948:114777, 2019.

\bibitem{Alshal:2019hpk}
H.~Alshal and T.~L. Curtright.
\newblock {Massive Dual Gravity in N Spacetime Dimensions}.
\newblock {\em JHEP}, 09:063, 2019.

\bibitem{Dawson:1982sc}
S.~Dawson and A.~N. Schellekens.
\newblock {Monopole Catalysis of Proton Decay in SO(10) Grand Unified Models}.
\newblock {\em Phys. Rev.}, D27:2119, 1983.

\bibitem{Georgi:1974sy}
H.~Georgi and S.~L. Glashow.
\newblock {Unity of All Elementary Particle Forces}.
\newblock {\em Phys. Rev. Lett.}, 32:438--441, 1974.

\bibitem{Arnold:1987mh}
Peter~Brockway Arnold and Larry~D. McLerran.
\newblock {Sphalerons, Small Fluctuations and Baryon Number Violation in
  Electroweak Theory}.
\newblock {\em Phys. Rev.}, D36:581, 1987.

\bibitem{Nath:2006ut}
Pran Nath and Pavel Fileviez~Perez.
\newblock {Proton stability in grand unified theories, in strings and in
  branes}.
\newblock {\em Phys. Rept.}, 441:191--317, 2007.

\bibitem{Baez:2009dj}
John~C. Baez and John Huerta.
\newblock {The Algebra of Grand Unified Theories}.
\newblock {\em Bull. Am. Math. Soc.}, 47:483--552, 2010.

\bibitem{Dorsner:2012nq}
Ilja Dorsner, Svjetlana Fajfer, and Nejc Kosnik.
\newblock {Heavy and light scalar leptoquarks in proton decay}.
\newblock {\em Phys. Rev.}, D86:015013, 2012.

\bibitem{Aaij:2014ora}
Roel Aaij et~al.
\newblock {Test of lepton universality using $B^{+}\rightarrow
  K^{+}\ell^{+}\ell^{-}$ decays}.
\newblock {\em Phys. Rev. Lett.}, 113:151601, 2014.

\bibitem{Alsaleh:2017ttv}
Salwa Alsaleh, Abeer Al-Modlej, and Ahmed Farag~Ali.
\newblock {Virtual black holes from the generalized uncertainty principle and
  proton decay}.
\newblock {\em EPL}, 118(5):50008, 2017.

\bibitem{Amati:1988tn}
D.~Amati, M.~Ciafaloni, and G.~Veneziano.
\newblock {Can Space-Time Be Probed Below the String Size?}
\newblock {\em Phys. Lett.}, B216:41--47, 1989.

\bibitem{Garay:1994en}
Luis~J. Garay.
\newblock {Quantum gravity and minimum length}.
\newblock {\em Int. J. Mod. Phys.}, A10:145--166, 1995.

\bibitem{Kempf:1994su}
Achim Kempf, Gianpiero Mangano, and Robert~B. Mann.
\newblock {Hilbert space representation of the minimal length uncertainty
  relation}.
\newblock {\em Phys. Rev.}, D52:1108--1118, 1995.

\bibitem{Adler:2001vs}
Ronald~J. Adler, Pisin Chen, and David~I. Santiago.
\newblock {The Generalized uncertainty principle and black hole remnants}.
\newblock {\em Gen. Rel. Grav.}, 33:2101--2108, 2001.

\bibitem{Ali:2009zq}
Ahmed~Farag Ali, Saurya Das, and Elias~C. Vagenas.
\newblock {Discreteness of Space from the Generalized Uncertainty Principle}.
\newblock {\em Phys. Lett.}, B678:497--499, 2009.

\bibitem{Vagenas:2018zoz}
Elias~C. Vagenas, Salwa~M. Alsaleh, and Ahmed Farag.
\newblock {GUP parameter and black hole temperature}.
\newblock {\em EPL}, 120(4):40001, 2017.

\bibitem{Vagenas:2018pez}
Elias~C. Vagenas, Ahmed Farag~Ali, and Hassan Alshal.
\newblock {GUP and the no-cloning theorem}.
\newblock {\em Eur. Phys. J.}, C79(3):276, 2019.

\bibitem{Vagenas:2019wzd}
Elias~C. Vagenas, Ahmed~Farag Ali, Mohammed Hemeda, and Hassan Alshal.
\newblock {Linear and Quadratic GUP, Liouville Theorem, Cosmological Constant,
  and Brick Wall Entropy}.
\newblock {\em Eur. Phys. J.}, C79(5):398, 2019.

\bibitem{Vagenas:2019rai}
Elias~C. Vagenas, Ahmed~Farag Ali, and Hassan Alshal.
\newblock {Massless charged particles, naked singularity, and GUP in
  Reissner-Nordström-de Sitter-like spacetime}.
\newblock {\em Phys. Rev.}, D99(8):084013, 2019.

\bibitem{Tejeiro:2010gu}
Juan~Manuel Tejeiro and Alexis Larranaga.
\newblock {Noncommutative Geometry Inspired Rotating Black Hole in Three
  Dimensions}.
\newblock {\em Pramana}, 78:155--164, 2012.

\bibitem{Abramowitz:1974:HMF:1098650}
Milton Abramowitz.
\newblock {\em Handbook of Mathematical Functions, With Formulas, Graphs, and
  Mathematical Tables,}.
\newblock Dover Publications, Inc., New York, NY, USA, 1974.

\bibitem{Damgaard:1988nq}
P.~H. Damgaard and H.~Huffel, editors.
\newblock {\em {Stochastic Quantization}}.
\newblock 1988.

\bibitem{Moffat:1996fu}
J.~W. Moffat.
\newblock {Stochastic gravity}.
\newblock {\em Phys. Rev.}, D56:6264--6277, 1997.

\bibitem{Moffat:2014eua}
J.~W. Moffat.
\newblock {Stochastic Quantum Gravity, Gravitational Collapse and Grey Holes}.
\newblock 2014.

\bibitem{Abazov:2008kp}
V.~M. Abazov et~al.
\newblock {Search for large extra dimensions via single photon plus missing
  energy final states at $\sqrt{s}$ = 1.96-TeV}.
\newblock {\em Phys. Rev. Lett.}, 101:011601, 2008.

\bibitem{Gingrich:2006hm}
Douglas~M. Gingrich.
\newblock {Black hole cross-section at the large hadron collider}.
\newblock {\em Int. J. Mod. Phys.}, A21:6653--6676, 2006.

\bibitem{Aaltonen:2008hh}
T.~Aaltonen et~al.
\newblock {Search for large extra dimensions in final states containing one
  photon or jet and large missing transverse energy produced in $p \bar{p}$
  collisions at $\sqrt{s}$ = 1.96-TeV}.
\newblock {\em Phys. Rev. Lett.}, 101:181602, 2008.

\bibitem{Khachatryan:2010wx}
Vardan Khachatryan et~al.
\newblock {Search for Microscopic Black Hole Signatures at the Large Hadron
  Collider}.
\newblock {\em Phys. Lett.}, B697:434--453, 2011.

\bibitem{Ali:2012mt}
Ahmed~Farag Ali.
\newblock {No Existence of Black Holes at LHC Due to Minimal Length in Quantum
  Gravity}.
\newblock {\em JHEP}, 09:067, 2012.

\bibitem{Cavaglia:2003qk}
Marco Cavaglia, Saurya Das, and Roy Maartens.
\newblock {Will we observe black holes at LHC?}
\newblock {\em Class. Quant. Grav.}, 20:L205--L212, 2003.

\end{thebibliography}
\bibliographystyle{unsrt}

\end{document}